\documentclass[prl,preprint,groupedaddress]{revtex4-1}

\usepackage{graphicx}
\usepackage{epsfig,amsmath,subfigure}
\usepackage{tikz}
\usepackage{color} \usepackage{amssymb} \usepackage{cancel}

\def\be{\begin{equation}}
\def\ee{\end{equation}}
\def\bea{\begin{eqnarray}}
\def\eea{\end{eqnarray}}
\def\beaa{\begin{eqnarray*}}
\def\eeaa{\end{eqnarray*}}

\def\f{\frac}

\def\befi{\begin{figure}}
\def\eefi{\end{figure}}

\def\bce{\begin{center}}
\def\ece{\end{center}}

\def\3int{\int\!\!\!\int\!\!\!\int}
\def\2int{\int\!\!\!\int}

\def\i {\textrm{i}} 
\def\e{\textrm{e}}

\def\black{\textcolor{black}}

\begin{document}


\title{Multiple-location matched approximation for Bessel function $J_0$ and its derivatives}


\author{Usama Kadri}
\affiliation 
{ 
$^{1}$School of Mathematics, Cardiff University, Cardiff CF24 4AG, UK\\
$^{2}$Department of Mathematics, Massachusetts Institute of Technology, Cambridge, MA, USA \\}

\date{\today}

\begin{abstract}
I present an approximation of Bessel function $J_0(r)$ of the first kind for small arguments near the origin. The approximation comprises a simple cosine function that is matched with $J_0(r)$ at $r=\pi/\e$. A second matching is then carried out with the standard, but slightly modified, far-field approximation for $J_0(r)$, such that first and second derivatives are also considered. The approximation is practical when nonlinear dynamics come into play, in particular in the case of nonlinear interactions that involve second order differential equations as in acoustic--gravity wave theory. A demonstration of the proposed matching technique applied to three-dimensional acoustic--gravity wave triad resonance in cylindrical coordinates is provided. 
\end{abstract}


\maketitle
\section{1. Introduction}
The boundary-value problem of the generation of gravity and/or acoustic waves in fluids due to a localised disturbance at the surface is associated with the three dimensional wave equation in cylindrical coordinates. Under standard conditions, the solution is given by Bessel functions which may become cumbersome for many problems, in particular when nonlinearity comes into play, such as in acoustic--gravity wave triad resonance \cite{longuet1950theory, kadri2013generation, kadri2015wave, kadri2016triad, kadri2016resonant}. Here we are concerned with an approximated solution valid within a distance of a few wavelengths from the disturbance origin, after which the waves are effectively damped or dissipated. Yet, for brevity, we do not consider damping or dissipation which can be easily added or treated numerically once a closed form solution is structured. {Fast and accurate Bessel function computations were presented in the literature, e.g. Ref. \cite{harrison2009}, though the nature of terms collected in the nonlinear problem discussed here require a much more simplified approach. Thus,} we present a simple cosine approximation for the near-field that is matched with $J(x)$ at $x=\pi/\e$. The near-field approximation is matched with a modified far-field approximation at various locations, in $x$, that take into account the first and second derivatives. This allows no singularities at the origin with incremental deviation from the exact solution in the near-field, and only small errors in the far-field. 

Since this work is originally motivated by the three-dimensional triad resonance of acoustic--gravity waves, the formulation of the problem and the general leading order solution are presented in the following section 2. The matched approximation and the validation of the proposed technique are given in sections 3 and 4, respectively. Then, an application example for a triad resonance is demonstrated in section 5, followed by concluding remarks in section 6.

\section{2. Formulation and solution}
Based on irrotationality, the general problem of gravity and/or acoustic waves is formulated in terms of the velocity potential $\varphi(r,z,t)$, where $\mathbf{u}=\nabla\varphi$ is the velocity field. We shall use dimensionless variables, employing $\mu h$ as lengthscale and $h/c$ as timescale. Assuming radial motion, the governing equation is then the compressible wave equation in cylindrical coordinates

\begin{equation} \label{eq:field}
\varphi_{tt}-\frac{1}{\mu^2}\left(\varphi_{rr}+\f{1}{r}\varphi_r+\varphi_{zz}\right)+\varphi_z=0.
\end{equation}
We seek modes that radiate with wavenumber $k$ and frequency $\omega$ in the form

\be \label{eq:sov}
\varphi=\mathbb{R}(r)\mathbb{Z}(z)\exp\left(\frac{1}{2}\mu^2z\right)\exp\left(-\i\omega t\right).
\ee
Upon substituting (\ref{eq:sov}) into (\ref{eq:field}), this separation of variables produces a set of two ordinary differential equations

\be \label{eq:diff}
\frac{\mathbb{R}_{rr}+\f{1}{r}\mathbb{R}_{r}}{\mathbb{R}}=-\f{\mathbb{Z}_{zz}+\left(\mu^2\omega^2-\f{1}{4}\mu^4\right)\mathbb{Z}}{\mathbb{Z}}=-k^2,
\ee
where $k^2$ is a separation constant between $r$ and $z$. The solution of the second ordinary differential equation is simple and can be found in Ref. \cite{kadri2016resonant}. 
%
%
The general solution of the first ordinary equation 
\be\label{eq:ode_R}
{\mathbb{R}_{rr}+\f{1}{r}\mathbb{R}_{r}}+k^2{\mathbb{R}}=0
\ee
is well known and it is given by Ref. \cite{abramowitz1964handbook}:
\be \label{eq:R_sol}
\mathbb{R}=C_1J_0(kr)+C_2Y_0(kr),
\ee
where $J_0$ and $Y_0$ are Bessel functions, and $C_1$, $C_2$ are coefficients that can be found from the space-time boundary conditions. Specifically, $J_0$ and $Y_0$ take the forms
\begin{eqnarray} \label{Bessel}
&&J_0(kr)=\sum_{n=0}^\infty\dfrac{\left(-k^2r^2/4\right)^n}{(n!)^2}\,,\\ &&Y_0(kr)=\frac{2}{\pi}\ln\left(\frac{kr}{2}\right)J_0(kr)-\frac{2}{\pi}\sum_{n=1}^\infty\left(\sum_{j=1}^n\frac{1}{j}\right)\dfrac{\left(-k^2r^2/4\right)^n}{(n!)^2},
\end{eqnarray}
In the far field, both $J_0$ and $Y_0$ oscillate and behave like damped cosine and sine functions, namely,
\begin{eqnarray}
J_0(kr)\approx\sqrt{\frac{2}{\pi kr}}\cos\left(kr-\frac{\pi}{4}\right)\,,\quad Y_0(kr)\approx\sqrt{\frac{2}{\pi kr}}\sin\left(kr-\frac{\pi}{4}\right)\ .
\end{eqnarray}
To avoid singularity at the origin we consider the case $Y_0=0$.

\section{3. Matched approximation}
The matched approximation comprises three steps. The \textbf{first step} of the matching is approximating the near-field using a simple cosine function of the form,
\be J_{0,near}(r) = \cos(ar),\ee where $a$ is a matching parameter calculated by matching $J_{0,near}$ with $J_0$ at $r=\pi/\e$. Thus, the approximation is exact at the origin and at the matching point $a\simeq 0.6967398$.

In the \textbf{second step} we match the near-field with a modified far-field approximation of the form,
\be J_{0,far}(r)=\sqrt{\frac{2}{\pi b_1 r}}\cos\left(b_2 r-b_3\right)\ee
where $b_1$, $b_2$ and $b_3$ are matching parameters. Obviously, when $b_1=b_2=1$ and $b_3=\pi/4$ the standard far-field approximation is met. However, these would result in a discontinuities with the near-field solution. Thus, instead of using the standard approximation we introduce a three-point matching, at $r=r_1, r_2, r_3$, such that the first and second derivatives are also matched. Thus, for the zeroth derivative we require that
\be
-\cos( a r_{11}) +\sqrt{\frac{2}{\pi b_1 r_{11}}}\cos\left(b_2 r_{11}-b_3\right) = 0,
\ee
and similarly for the second and third derivatives,
\be
a\sin(ar_{12})-b_2\sqrt{\f{2}{\pi b_1r_{12}}}\sin(b_2r_{12}-b_3)-\f{1}{2}\sqrt{\f{2}{\pi b_1r_{12}^3}}\cos(b_2r_{12}-b_3)=0,
\ee
\be
a^2\cos(ar_{13})-\left(\f{b_2^2}{\sqrt{r_{13}}}-\f{3}{4\sqrt{r_{13}^5}}\right)\sqrt{\f{2}{\pi b_1}}\cos(b_2r_{13}-b_3)=0.
\ee
The choice of $r_{11}$, $r_{12}$, $r_{13}$ is made where the near-field solution is in the neighbourhood of the exact solution of $J_0$, though not necessarily identical, e.g. $r_{11}=0.698132$, $r_{12} = 0.926421$,  $r_{13}=3.47390$. With these numerical values, we have a set of three equations with three unknowns, that give $b_1=1.09713$, $b_2=1.03946$, $b_3=0.972672$. The third step of the approximation is described below.

\section{4. Numerical Validation}
It is easy to see that the proposed approximation is exact at the origin and at the matching, $x=\pi/\e$, whereas (\ref{eq:ode_R}) is satisfied with an error negligible in the near-field and below $\% 3$ for the standard case of $b_1=b_2$, and $b_3=\pi/4$. For the other combinations of $b_j$ the error  in (\ref{eq:ode_R}) is again negligible in the near-field, and below $\%8$ elsewhere as illustrated in figure 1. The matching at $\pi/\e$ allows almost an exact approximation of the near-field and a smooth transition to the far-field as illustrated in figure 1. However, this result can only be useful if the derivatives are not concerned, since they cannot be smooth. To obtain a proper approximation for the derivatives as well, we employ the multiple-location matching described above. Taking into account $r_{11}$, $r_{12}$, $r_{13}$, $b_1$, $b_2$, and $b_3$ presented in section 4 we are able to design a solution whose first and second derivatives are also matched, see figure 2. Since matching the far-field \textit{deforms} the standard far-field solution of $\mathbb{R}$, the solution diverges as $r\to\infty$, and a clear discrepancy is noticed after a few periods only. To overcome this difficulty, we impose a \textbf{third step} with a second three-location far-field matching at $r_{21}$, $r_{22}$, $r_{23}$, where the standard far-field parameters, $b_1=b_2=1$, $b_3=\pi/4$, and $\mathbb{R}$ and its derivatives are matched exactly. Here, $r_{2j}$ $(j=1,2,3)$ are chosen points where $J_{0,far}$ with $r_{1j}$ match (or tangent to) the exact solution. This guarantees that the second far-field approximation and derivatives tend to the exact solution its derivatives for large $r$, as illustrated in figure 3.
\begin{figure} [h!] \label{fig:approx_err}
\centering{ \epsfig{figure=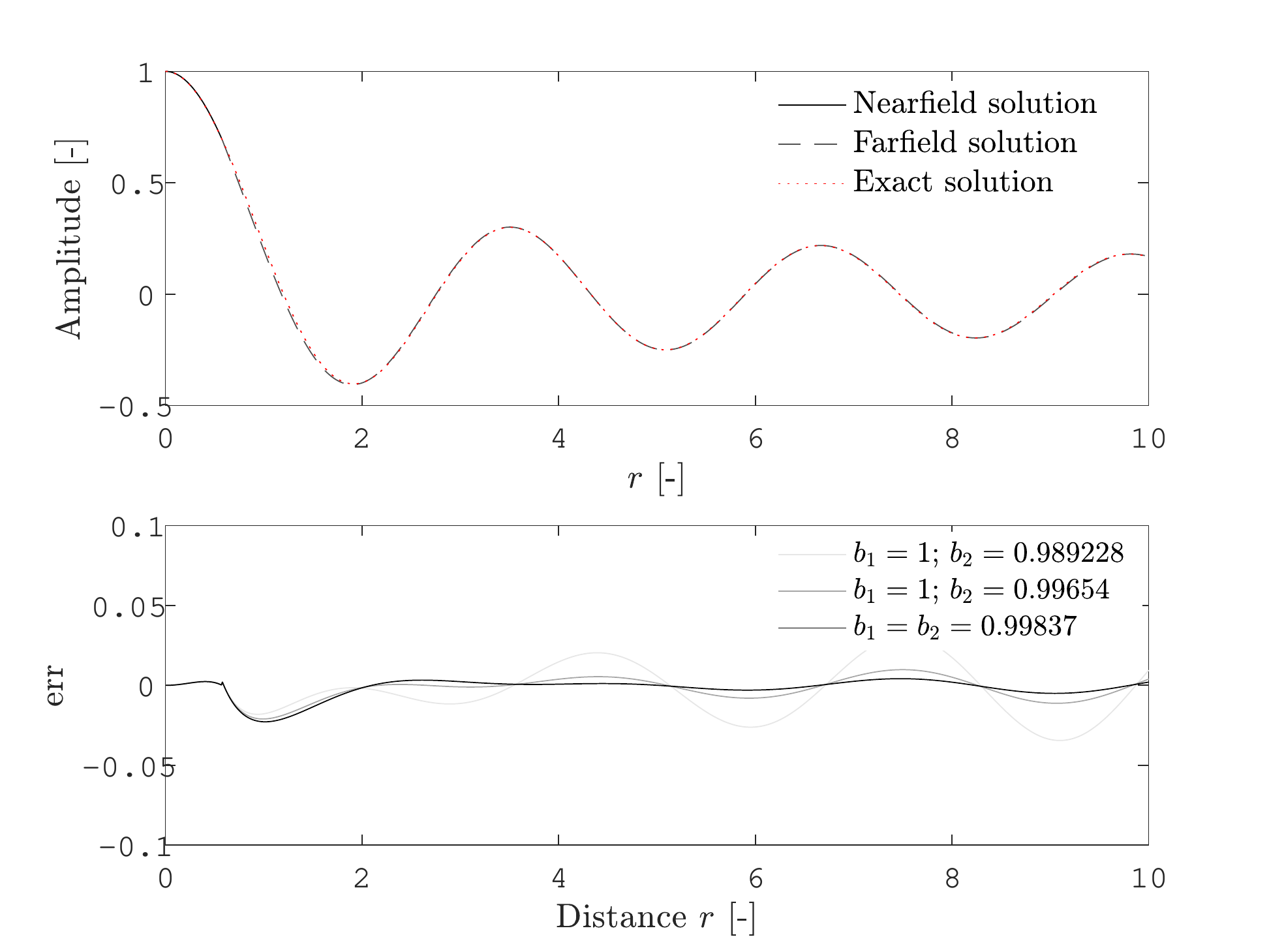,width=12cm}}
\caption{Top: Exact (dotted), near-field (solid) and far-field (dashed) solutions of $\mathbb{R}$. Bottom: normalised error in the approximate solution for different matching parameters $b_1$ and $b_2$.}
\end{figure}
\begin{figure} [h!] \label{fig:approx_I} 
\centering{ \epsfig{figure=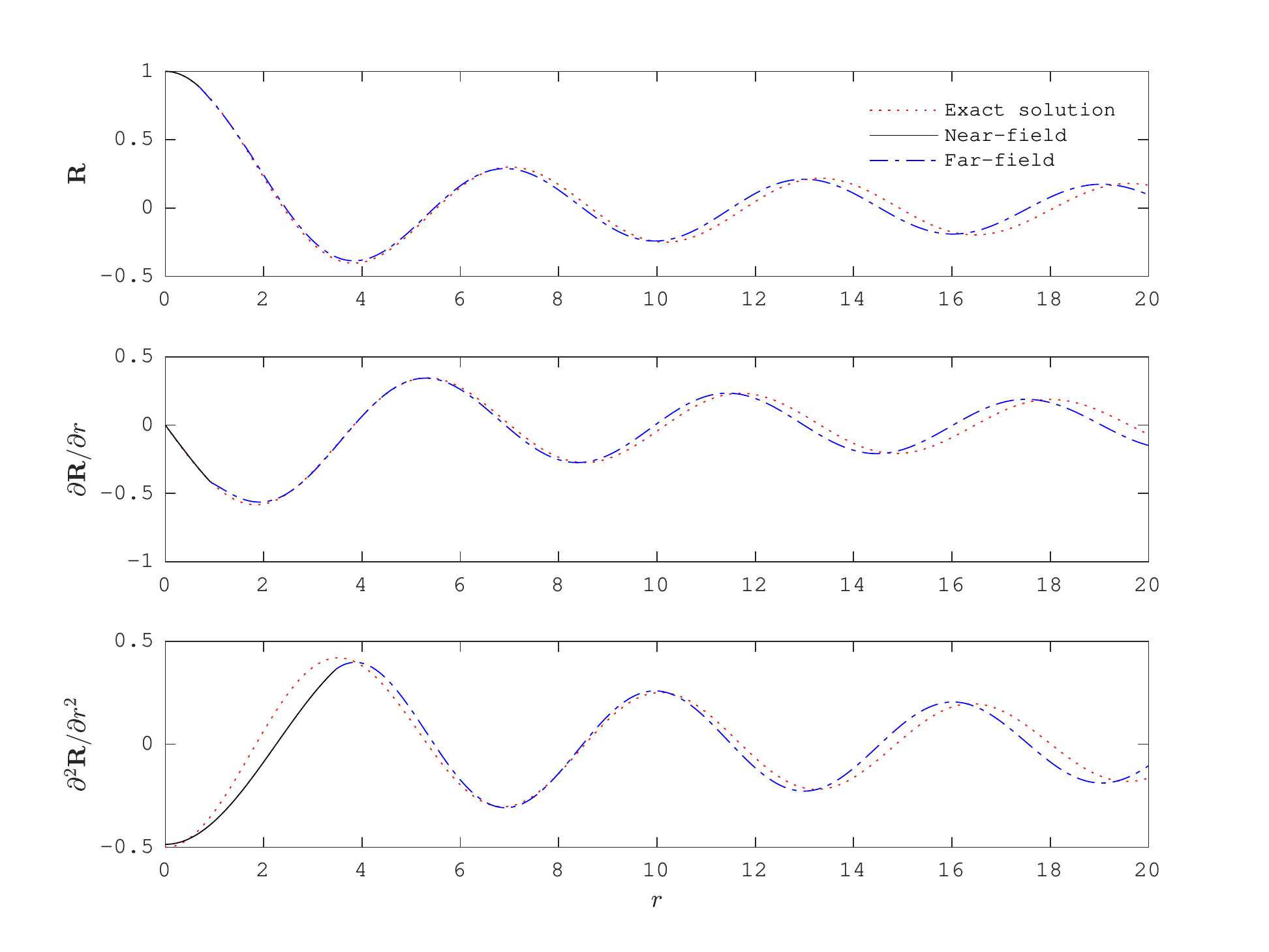,width=12cm}}
\caption{Exact (dotted), near-field (solid) and far-field (dashdotted) solutions of $\mathbb{R}$ and its first two derivatives. Here, $r_{11}=0.69813$, $r_{12}=0.92642$, $r_{13} = 3.4739$.}
\end{figure}
\begin{figure} [h!] 
\centering{ \epsfig{figure=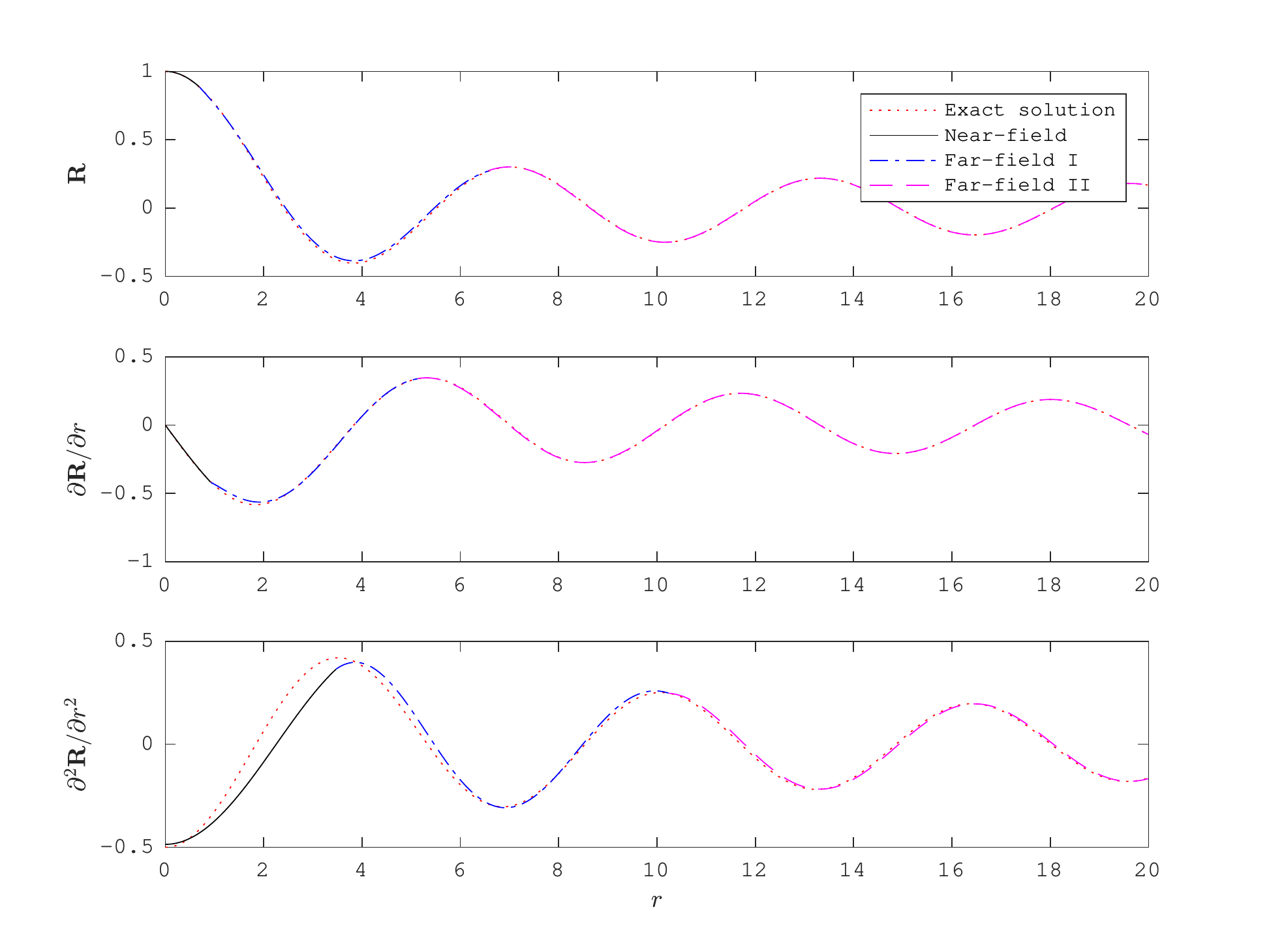,width=12cm}}
\caption{Exact (dotted), near-field (solid), first far-field (dashdotted), and second far-field (dashed) solutions of $\mathbb{R}$ and its first two derivatives. Here, $r_{21} = 6.5845$, $r_{22}= 5.1403$, $r_{32}=10.241$.}
\end{figure}

\section{5. Nonlinear triad interaction: application example}
Consider the cubic nonlinear acoustic--gravity wave equation
\begin{equation} \label{eq:field}
\varphi_{tt}-\frac{1}{\mu^2}\left(\varphi_{rr}+\f{1}{r}\varphi_r+\varphi_{zz}\right)+\varphi_z+|\nabla\varphi|^2_t+\tfrac{1}{2}\nabla\varphi\cdot\nabla\left(|\nabla\varphi|^2\right)=0\,,
\end{equation}
with a velocity potential for three interacting modes, 
\begin{eqnarray} \label{eq:potential_faraday}
\varphi&=&\epsilon \left\{S_+(T)\mathbb{R}(k_+r)\e^{|k_+|z}\black{\e^{-\i\omega_+ t}}+\text{c.c.}\right\}+\epsilon \left\{S_-(T)\mathbb{R}(-k_-r)\e^{|k_-|z}\black{\e^{-\i\omega_- t}}+\text{c.c.}\right\}\nonumber\\
&&+\alpha\left\{A(T,R)\black{\mathbb{R}(\mu\kappa r)}\cos\omega_n(Z+1)\black{\e^{-\i\omega t}}+\text{c.c.}\right\}+ \dots.
\end{eqnarray} 
To form a resonant triad the dispersion relations for the acoustic and gravity waves need to be satisfied, see (2.8) and (2.12) in Ref.\cite{kadri2016resonant}. The first two terms in (\ref{eq:potential_faraday}) represent the two surface gravity waves while the last represents the acoustic mode, that is scaled in the vertical coordinate $Z=\mu z$. The surface wave amplitudes $S_{\pm}$ and the acoustic mode amplitude $A$ depend on the `slow' time $T=\mu t$, and space $R=\mu r$, where $\epsilon=\alpha\mu^{1/2}$ with $\alpha=O(1)$  (see Ref. \cite{kadri2016resonant}). In addition, these modes must obey the resonance conditions
 \be  \textrm{(i) } k_++k_-=\mu\kappa; \qquad \textrm{(ii) } \omega_++\omega_-=\omega,\ee
where $k_{\pm}$ are the gravity wave numbers, and $\mu\kappa$ is the acoustic wavenumber.

Terms in $\mathbb{R}$ cause secular behaviour at higher order expansions, and require imposing solvability conditions that lead to the amplitude evolution equations. Considering the exact solution given in \ref{eq:R_sol}, would require using Bessel function of the first kind and its first and second spatial derivatives, 
\be \label{eq:R} \mathbb{R} = J_0;\quad \mathbb{R}_r = -J_1; \quad \mathbb{R}_{rr} = -\frac{1}{2}\left(J_0-J_2\right). \ee
In particular, the terms $\varphi_r$, $\varphi_{rr}$, $\varphi_r^2$, $|\nabla^2\varphi|_t$, and  $\tfrac{1}{2}\nabla\varphi\cdot\nabla\left(|\nabla\varphi|^2\right)$ of (\ref{eq:field}) after substituting (\ref{eq:R}) in (\ref{eq:potential_faraday}), make collecting resonant terms and seeking solvability conditions rather cumbersome, if at all possible. 

To carry out with the nonlinear settings we instead apply the proposed multiple-location matching, which gives,
\be \label{eq:R_approx} \mathbb{R} = \cos{ar};\quad \mathbb{R}_r = -a\sin(ar); \quad \mathbb{R}_{rr} = -a^2\cos(ar), \ee
for the near-field, and 
\be \begin{split} \label{eq:R_approx} \mathbb{R} =\sqrt{\frac{2}{\pi b_1 r}}\cos\left(b_2 r-b_3\right); \\ \mathbb{R}_r = -\sqrt{\f{2}{\pi b_1r_{12}}}\left(b_2\sin(b_2r_{12}-b_3)+\f{1}{2r_{12}^3}\cos(b_2r_{12}-b_3)\right);\\ 
 \mathbb{R}_{rr} = -\left({b_2^2}-\f{3}{4{r_{13}^2}}\right)\sqrt{\f{2}{\pi b_1r_{13}}}\cos(b_2r_{13}-b_3), \end{split}\ee
for the far-field. Thus, when substituting (\ref{eq:potential_faraday}) in the governing equation, we focus on terms $\propto$ $\exp\{\textrm{i}(ak_{\pm}r_{\pm}-\omega_{\pm} t)\}$, and $\exp\{\textrm{i}(a\kappa r-\omega t)\}$, which are much more convenient to work with. For example, in the near-field collected terms $\propto \exp\{\textrm{i}(a\mu\kappa r-\omega t)\}$ are:
\[\varphi_r \to \alpha\i a \mu\kappa A \cos\omega_n(Z+1) ; \qquad \varphi_{rr} \to -\alpha a^2 \mu^2\kappa^2 A \cos\omega_n(Z+1)\]
\[\varphi^2_r \to 2\epsilon^2a^2k^2S_+S_-\e^{2kz}; \qquad |\nabla\varphi|^2_t=\left(\varphi^2_r+\varphi_z^2\right)_t\to -2\i\omega\epsilon^2S_+S_-\e^{2kz}k^2(a^2+1)\]
%
Similarly, the far-field terms can be collected, and following a similar procedure as in Ref. \cite{kadri2016resonant} the desired evolution equations can be obtained. 

\section{Discussion}
We present a simple near-field approximation for Bessel function of the first kind, $J_0$. The approximation comprises a simple cosine function with an argument that is matched with $J_0(r)$ at exactly $r=\pi/\e$ resulting in a highly accurate approximation. This leads to two main straight forward questions: what makes a cosine a good approximation function? and what is special about $\pi/\e$? The answer to the first question is rather straight forward, as expanding $\cos{ar}$ into a series one can conclude that for small arguments the leading terms are very similar to their counterpart in the Bessel function of the first kind. The second question remains open. 

For the far-field, we employ the standard far-field approximation. However, this cannot be smooth with our solution, or its derivatives. Therefore, we introduce an \textit{intermediate zone} that we refer to as far-field I, whereby the standard approximation is modified with three more parameters, that are matched with the Bessel corresponding derivatives, from one hand, and three different locations that allow matching with the near-field, on the other hand. Since, such modification creates some discrepancy in the actual far-field solution, referred to as far-field II, we allow a second matching with the standard far-field, at the three proper locations in space. 

The proposed approximation deals with first two derivatives only, though in principle one can introduce a Polynomial, in $b_{ij}$, of enough orders to match as many derivatives as required. Finally, the proposed approximation was motivated from the nonlinear interaction of acoustic--gravity waves in cylindrical coordinates. As demonstrated briefly, the approximation makes collecting resonant terms straight forward avoiding a cumbersome work that involves Bessel function derivatives and their second and third powers. 

\end{document}